# Coherent Laser Induced Synthesis of Rare Earth Doped Nanocrystallites of $50PbO–25Bi_2O_3–20Ga_2O_3–5BaO$

Iwan Kityk[1]

[1]Institute of Optoelectronics and Measuring Systems, Czestochowa University of Technology, PL-42201, Czestochowa, Poland
iwank74@gmail.com

Ahmed El-Naggar[2,3], Abdullah Albassam[2]

[2]Research Chair of Exploitation of Renewable Energy Applications in Saudi Arabia, King Saud University, P.O. Box 2455, Riyadh 11451

[3]Physics Department, Ain Shams University, Abbasia, Cairo 11566, Egypt

Nazariy Andrushchak[4], Daniel Kulwas[5], Pawel Czaja[1], Bouchta Sahraoui[6]

[4]Lviv Polytechnic National University, 79013, Lviv, Ukraine

[5]Energia Oze Sp. z o.o. 42-274 Czestochowa, Poland

[6]University of Angers 49035, Anger Cedex 01, France

*Abstract*—A principal possibility of formation the nanostructures on the surfaces of $50PbO–25Bi_2O_3–20Ga_2O_3–5BaO$ (doped by $Eu^{3+}$, $Er^{3+}$, $Dy^{3+}$) is demonstrated by using multi-coherent beams. As a sources of the photoinducing coherent light we have used nanosecond Nd:YAG and Er:Yb lasers generating at 1064 nm and 1540 nm, respectively. The morphology of the photoinduced surfaces is sensitive to the type of rare earth ions. The thickness of the layer was about 20…30 nm. Possible mechanisms are explained by coherent photoinduction of the valence electrons.

*Keywords—nanolayers; photoinduced glasses; rare earth insert.*

## I. Introduction

A possibility to modify surface nanostructures under external laser light is a hot topic of modern nanotechnology.

Among the different materials the oxide glasses or ceramics possessing heavy elements and rare earths seem to be more promising [1]. Their specific features is high stability to the laser power densities in the visible and near infrared spectral range, good controlled technology of incorporation of the rare earths, high space distribution of the rare earths, high thermal stability etc. The modification of the surfaces under the one laser beam is not efficient; however, usage of the two coherent beams of the same lasers, which are incident at different angles, may be very promising way for their modifications and occurrence of nano-sized morphology. In this report we have modified the $50PbO–25Bi_2O_3–20Ga_2O_3–5BaO$ (doped by $Eu^{3+}$, $Er^{3+}$) that is demonstrated by using multi-coherent beams. As a sources of the photoinducing coherent light we have used Nd:YAG and Er:Yb nanosecond lasers generating at 1064 nm and 1540 nm, respectively. The morphology of the photoinduced surfaces is sensitive to the type of rare earth ions. It may open a new approach for the technology of formation the surface nanostructures, which is more flexible and chipper. Thus, we will need to create these components, incorporating the applicable criteria.

## II. Methodology

### A. Sample preparation

Initial materials of purity about 99.997% were mixed to yield 50 g batches of $50PbO–25Bi_2O_3–20Ga_2O_3–5BaO$ (mol%) composition. 99.996% $RE_2O_3$ (RE=$Eu^{3+}$, $Dy^{3+}$, $Er^{3+}$) were added to the starting powders to synthesize glasses doped with 0.5% of $RE_2O_3$. For melting at 1030 K for 25 min. it was used a platinum crucible. In order to prevent water contamination in the glasses we applied Ar gas flow at the rate 10 l/min. Quenching procedure was carried out in an air and the samples have been annealed at 580 K during 2 h. X-ray diffraction confirms an absence of crystallization. General procedure of their synthesis is described in [2, 3].

The specimens were in the form of a parallelepiped with sizes $8 \times 8 \times 1$ mm$^3$. The surfaces were polished in order to obtain surface roughness better than 14 μm. Optical transparency was measured by single-beam spectrophotometer and Fourier spectrometer within the spectral range 0.5–9 μm with spectral resolution about 1 cm$^{-1}$. The scattering background was below 0.3 % by intensity. The et-up- allowed to perform space scanning of the fundamental beams and detect their changes during a long time. All this equipment was calibrated by specimens with the known parameters of nonlinear optical constants.

## B. Laser induced setup

The laser coherent treatment (see Fig. 1) was performed by simultaneous coherent treatment of Er:Yb laser at 1540 nm and the Nd:YAG laser at 1064 nm with power densities permanently changing up to 120 MW/cm$^2$. The applied Er:Yb glass laser was exceptionally stable as a variable pulsed block possessing space distributed operation. The quantum generator was supplied by Cr doped passive Q switch and allowed to change the laser pulse duration time within 22 ns…310 ns and pulse frequency repletion was varied within the 110 Hz…850 Hz. The diameter of the beam was changed in the range 0.8…1.7 mm. Varying these parameters it was possible to vary the effective energy and power densities of the laser. The system of mirrors (M), polarizer (P), beam splitters (BS) form coherent beams with diameters up to 3 mm incident on the glass polished surfaces. Thus, such treated samples are explored by JEOL 2010 F TEM with resolution up to 5 nm.

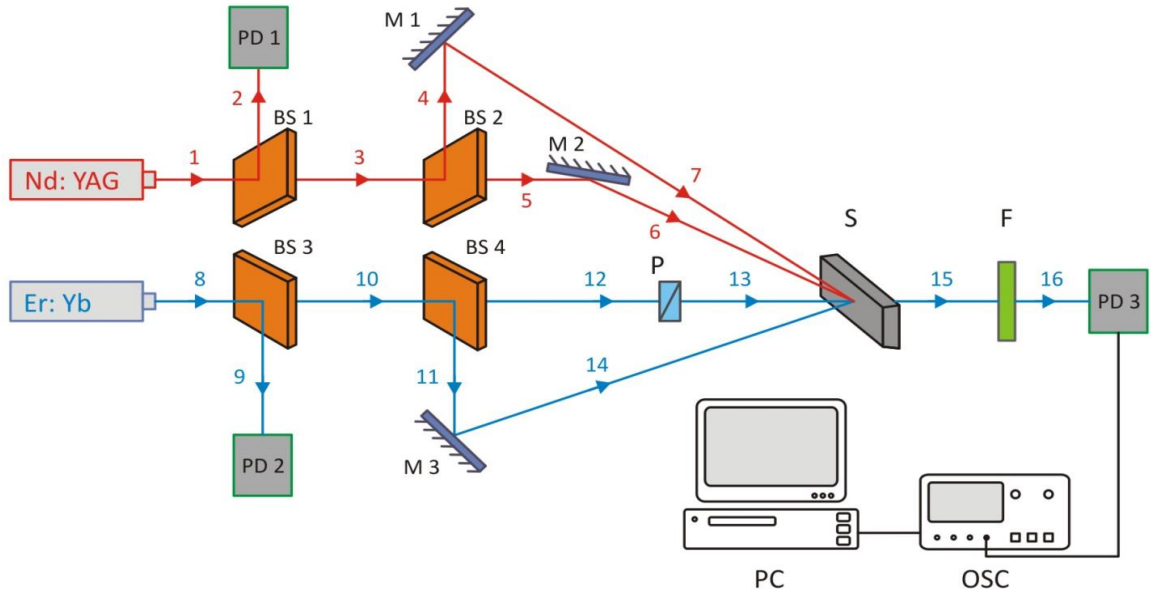

Fig. 1. Principal setup for multi-color coherent treatment

The control of the photoinduced intensities and the light harmonics were done by the set of photodetectors: PD1, PD2, PD3.

### III. RESULTS AND DISCUSSION

In the Fig. 2 the occurred laser modified surfaces for the studied glasses are presented. One can see that at the fixed power densities about 120 MW/cm$^2$ the TEM picture is principally different. The more clear nanopictures are observed for the Dy$^{3+}$ doped samples. For the two others of such formed morphology is more aggregate-like. The principal mechanisms are caused by the photoinduced excitations of the localized RE f-d levels and their interactions with the anharmonic phonons [4].

Depending on the radiuses of the RE ions we observe the different coordination of the surrounding oxygen-like ligands. The corresponding delocalized 2pO levels seem to be very sensitive to the external light.

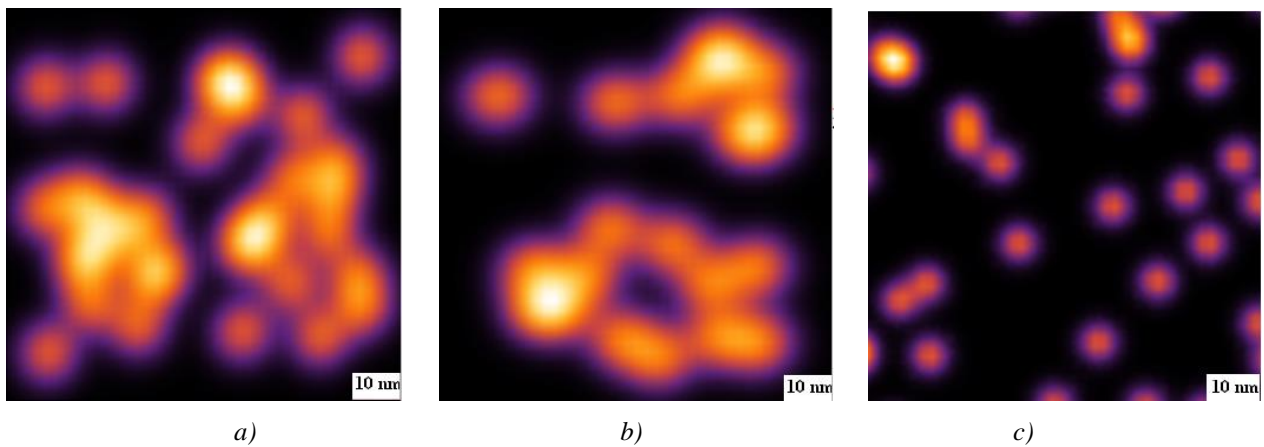

*a)*          *b)*          *c)*

Fig. 2. Laser induced structure during simultaneous coherent treatment by Er:Yb and YAG:Nd coherent lasers at efficient power densities about 120 MW/cm$^2$: doped by a) Eu$^{3+}$, b) Dy$^{3+}$, c) Er$^{3+}$.

It is crucial that applied one beam treatment did not caused the changes in the dependences. It may reflect an occurrence of gratings, which may play here crucial role. The appeared nano-surfaces are relatively strong and may exists up to 2-30 days. The increasing temperature destroy such formed gratings.

The control of the second and third harmonics have shown that the observed gratings are non-cetrosymmetrics because it was observed signals of the second harmonic generation which is described by third rank polar tensors. In order to separate the parasitic fluorescence background we have done the spectral studies of the reflected light in the spectral range up to 10 nm outside the SHG resonance at 770 nm. The SHG intensity was at lest one order higher that the background. The space meam map profile at wavelength 770 nm is shown in the Fig.3. The scattering light background did not exceed 3…5 %. The second harmonics signal was significantly higher than for the third harmonics.

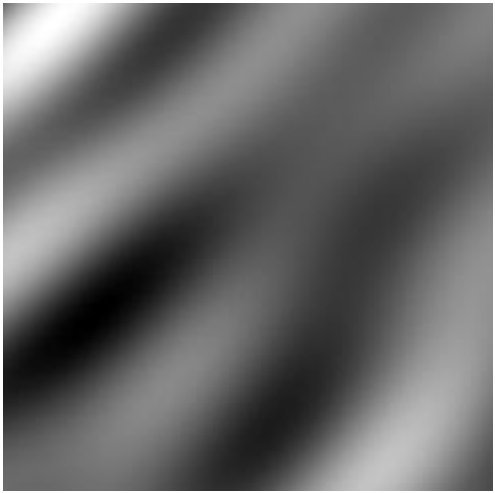

Fig. 3. Beam space distribution of the second harmonic signal at the doubled frequency wavelength of 770 nm

Following this profile one can see that the structure possess principal fragments of the gratings that was formed due to the laser induced treatment.

The TEM and AFM control have shown that the thickness of the laser induced gratings did not exceed 70 nm and they possess some space gradients extending intra the deeper levels. The surface non-homogeneity indicates that the non-homogeneity did not exceed 5 %. Therefore, it may reflect the different phase matching conditions in the different points of the laser treated surface. The optimal angles between the photoinduced beams for Nd:YAG were about 20…21 degree and for the Er:Yb lasers were 32…33 nm.

The polarization of eh laser induced light did not have any influence on the structure. The space distribution of the beams also is not crucial here.

The observed data allow to form the laser operated setup which may also change the effective generation of the laser beam. The such kind of devices may be very useful for self-forming nanostructures which possess efficient second order nonlinear optical effects described by the third rank polar tensors. At the same time, the phase matching conditions allow to achieve additional increase of the output nonlinear optics. Because the principal origin of the effect is connected with phonon anharmonicities one can expect an enhancement of the effects with temperature.

We have done studies in the wide temperature range (see Fig. 4) which shows high sensitivity to temperature.

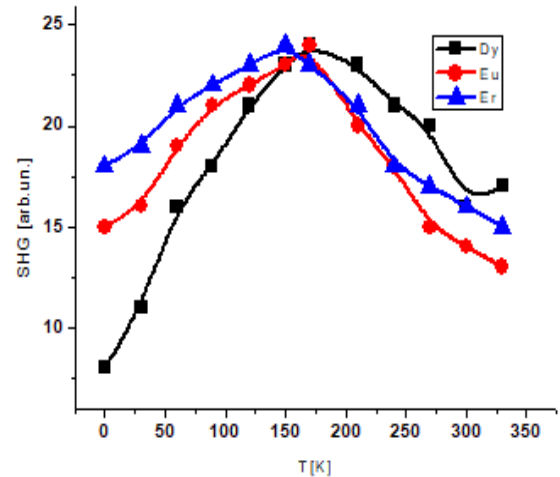

Fig. 4. Temperature dependences of the SHG for different doped glasses

Following the Fig. 4 it is clear that the temperature dependences show both with respect to the position of the maximum as well as well as by the shape.

The principal mechanisms may be explained by significant influence of temperature of the local environment and change of photopolarization [5, 6]. Particular interest present Er doped materials, which have the SHG maximum at the lowest temperature (about 150 K). The obvious asymmetry may be caused by different phonon contributions at different temperatures. Additionally $Er^{3+}$ ions possess strong absorption cross section in the spectral range 500-700 nm [7-10] $Er^{3+}$-doped materials possess very principal high luminescence quenching concentration compared to other rare-earth ions. Hence these glasses cover S-band extending from 1460 to 1530 nm, C-band (1530- 1565 nm), L- bands (1565- 1625nm) and part of the U- band (1625- 1675 nm). Moreover these glasses exhibits green emission under excitation wavelength at 445 nm for transition $^2H_{11/2} \rightarrow {}^4I_{15/2}$ (528 nm) and $^4S_{3/2} \rightarrow {}^4I_{15/2}$ (545nm). The role of phonon contributions is caused preliminary by anahrmonic phonons.

The varied rare ions possess different photo-excitations cross-sections. Consequently, the photo-mechanical changes also may be different. Appropriately varying the polarizabilities of the cations in the basic glasses one can operate by their properties. Moreover, the coherent laser treatment may also cause local photo-structural changes including the photo-crystallization. Therefore, we deal with the samples similar to the ceramics. The photo-excited transport between the localized rare earth ions and delocalized p-states may be very important for such kind of materials.

The further efforts will be devoted to find optimal concentrations of the rare earths and the appropriate intensities of the particular coherent wavelengths. This ratio may be principal for the photo induced grating formation. Consequently, the variation by the nanoparticle sizes and shapes is very crucial for optoelectronics [11-19]. The latter may enhance photo polarization which in turn change the resonance conditions of the SHG.

The future work will be devoted to a search of glass materials with the varied heavy cations and the anions, which determine laser operation efficiency.

## IV. CONCLUSIONS

For the first time it was shown a possibility to form the near the surface nanostructured morphology using external laser treatment. $50PbO–25Bi_2O_3–20Ga_2O_3–5BaO$ glasses (doped by $Eu^{3+}$, $Er^{3+}$, $Dy^{3+}$ were treated by using multi-coherent beams. As a sources of the photoinducing coherent light we have used nanosecond Nd:YAG and Er:Yb lasers generating at 1064 nm and 1540 nm wavelengths, respectively. The morphology of the photoinduced surfaces is sensitive to the type of rare earth ions. It was established that only the two beam coherent treatment performed by simultaneously by Er:Yb laser at 1540 nm and the Nd:YAG laser at 1064 nm with power densities permanently changing up to 120 MW/cm$^2$. The signals of the SHG are similar to the gratings like structure. The real laser induced nanostructure exists during the long period, up to several months and may be considered like a beginning of a new technology for optical triggering which is crucial in the different laser deflectors and laser light modulators. Additionally the proposed set-up may serve as a sensors of high intensity light during preparation of the high power laser devices.

ACKNOWLEDGMENT

The authors are grateful to the Deanship of Scientific Research, King Saud University for funding through Vice Deanship of Scientific Research Chairs. Also, this result of investigation is a part of a project that has received funding from the European Union's Horizon 2020 research and innovation program under the Marie Skłodowska-Curie grant agreement No 778156.